\documentclass[10pt,aps,prb,twocolumn,preprintnumbers,amsmath,amssymb,showpacs,citeautoscript,superscriptaddress, floatfix]{revtex4-2}
\usepackage[latin1]{inputenc}
\usepackage[T1]{fontenc}
\usepackage[english]{babel}
\usepackage[pdftex]{graphicx}
\usepackage{tabularx}
\usepackage{amsmath}
\usepackage{times}
\usepackage[usenames,dvipsnames,svgnames,table]{xcolor}
\usepackage[colorlinks,plainpages=false,linkcolor=blue,urlcolor=blue,citecolor=blue,pdfpagemode=UseNone]{hyperref}
\usepackage{svg}

\renewcommand{\AA}{\text{\r{A}}}

\begin{document}

\title
{
\boldmath
Delafossites as an unexpected competing phase to infinite-layer oxides
}

\author{Armin Sahinovic}
\affiliation{Department of Physics and Center for Nanointegration (CENIDE), Universit\"at Duisburg-Essen, Lotharstr.~1, 47057 Duisburg, Germany}
\author{Benjamin Geisler}
\email{benjamin.geisler@ufl.edu}
\affiliation{Department of Physics and Center for Nanointegration (CENIDE), Universit\"at Duisburg-Essen, Lotharstr.~1, 47057 Duisburg, Germany}
\affiliation{Department of Physics, University of Florida, Gainesville, Florida 32611, USA}
\affiliation{Department of Materials Science and Engineering, University of Florida, Gainesville, Florida 32611, USA}
\author{Rossitza Pentcheva}
\email{rossitza.pentcheva@uni-due.de}
\affiliation{Department of Physics and Center for Nanointegration (CENIDE), Universit\"at Duisburg-Essen, Lotharstr.~1, 47057 Duisburg, Germany}

\date{\today}

\begin{abstract}
Motivated by the discovery of superconductivity in Sr-doped infinite-layer nickelate films on SrTiO$_3$(001),
we explore the broader landscape of $AB$O$_2$ oxides through comprehensive high-throughput first-principles simulations.
Specifically, delafossites and their ordered rock-salt (111) variants stand out
as intriguing layered oxides that share the infinite-layer $AB$O$_2$ stoichiometry and simultaneously retain a perovskite-like octahedral motif.
This positions them as a unique structural bridge between these two phases and as promising candidates for novel correlated electronic states. 
We compile a phase diagram 
that compares the relative stability of these four distinct oxides across the periodic table.
Surprisingly, we find that the delafossite structure rivals the infinite-layer phase in thermodynamic stability for the nickelates,
and even more for the recently suggested palladate and platinate analogs.
Comparison of the respective electronic structures reveals that the delafossite compounds, which we find to be characterized by reversed cation order,
exhibit a strongly $d_{z^2}$-dominated Fermi surface, in stark contrast to the $d_{x^2-y^2}$ character observed in the infinite-layer phases.
Among all candidates, the La-Ni combination stands out as a thermodynamic optimum for stabilizing the infinite-layer motif.
Furthermore, we show that hole doping via Ca, Sr, and Ba systematically enhances the stability of the infinite-layer phase in all three transition-metal families.
These results reveal fundamental challenges in realizing bulk substrate-free infinite-layer oxides,
and simultaneously offer guidance for future experimental synthesis efforts targeting novel superconducting compounds.


\end{abstract}

\maketitle

\section{Introduction}

Since the discovery of superconductivity in Sr-doped NdNiO$_2$ films grown on SrTiO$_3$(001)~\cite{Li-Supercond-Inf-NNO-STO:19, Li-Supercond-Dome-Inf-NNO-STO:20, Zeng-Inf-NNO:20},
infinite-layer ($AB$O$_2$) nickelates have attracted considerable interest~\cite{Nomura-Inf-NNO:19, JiangZhong-InfNickelates:19, Sakakibara:20, JiangBerciuSawatzky:19, Botana-Inf-Nickelates:19, Lechermann-Inf:20, Si-Zhonh-Held:InfNNO-Hydrogen:20, NNO-SC-Thomale:20, Gu-NNO2:20, Lu-MagExNdNiO2:21, SahinovicGeisler:21, Wang-IL-Pauli:21, Geisler-VO-LNOLAO:22, Zeng-Inf-NNO:22, KreiselLechermann-IL:22, Rossi-IL-CO:22, Fowlie-IL-IntrinsicMag:22, SahinovicGeisler:22, Wang-Pressure-PNO:22, SahinovicGeislerPentcheva:23, Geisler-Rashba-NNOSTOKTO:23},
as they constitute a long-sought-for class of $3d^9$ cuprate-like superconductors.
Subsequent work rapidly increased the family of superconducting nickelates, for instance, 
to PrNiO$_2$ and LaNiO$_2$ films~\cite{Osada-PrNiO2-SC:20, Osada-LaNiO2-SC:21}
as well as to quintuple Ruddlesden-Popper-derived films~\cite{Pan-ILSC:22}.
Early on, infinite-layer palladates and platinates have been suggested as isoelectronic alternatives to the nickelates~\cite{Kitatani-AritaZhongHeld:20}.
Theoretical investigations predicted a comparable critical temperature~\cite{kitatani_optimizing_2023} for these compounds,
yet a lower oxygen vacancy formation energy at the apical oxygen sites~\cite{SahinovicGeisler:21}.

In sharp contrast to recent reports of $T_c \sim 80$~K in pressurized La$_3$Ni$_2$O$_7$~\cite{Sun-327-Nickelate-SC:23, Hou-LNO327-ExpConfirm:23, Zhang-LNO327-ZeroResistance:23, Luo-LNO327:23, Geisler-LNO327-Structure:24, Geisler-LNO327-Optical:24},
superconductivity in a bulk infinite-layer nickelate remained elusive so far~\cite{Li-NoSCinBulkDopedNNO:19, Wang-NoSCinBulkDopedNNO:20, Hu_atomic_scale_disorder_bulk_NNO_24}
and is limited to film geometries.
This raised early questions about the role and composition of the polar interface in these systems~\cite{GeislerPentcheva-InfNNO:20, BernardiniCano:20, He-IL:20, Zhang-IL:20, GeislerPentcheva-NNOCCOSTO:21, GoodgeGeisler-NNO-IF:22}.
Recent progress towards free-standing infinite-layer nickelates still requires confinement by several layers of SrTiO$_3$~\cite{Lee-IL-Nickelates-Freestanding:24, Yan-IL-Nickelates-Freestanding:24}.
However, the absence of structural constraints by the substrate may promote the formation of competing phases, even with distinct symmetry. 

\begin{figure}[t]
    \includegraphics[width=\linewidth]{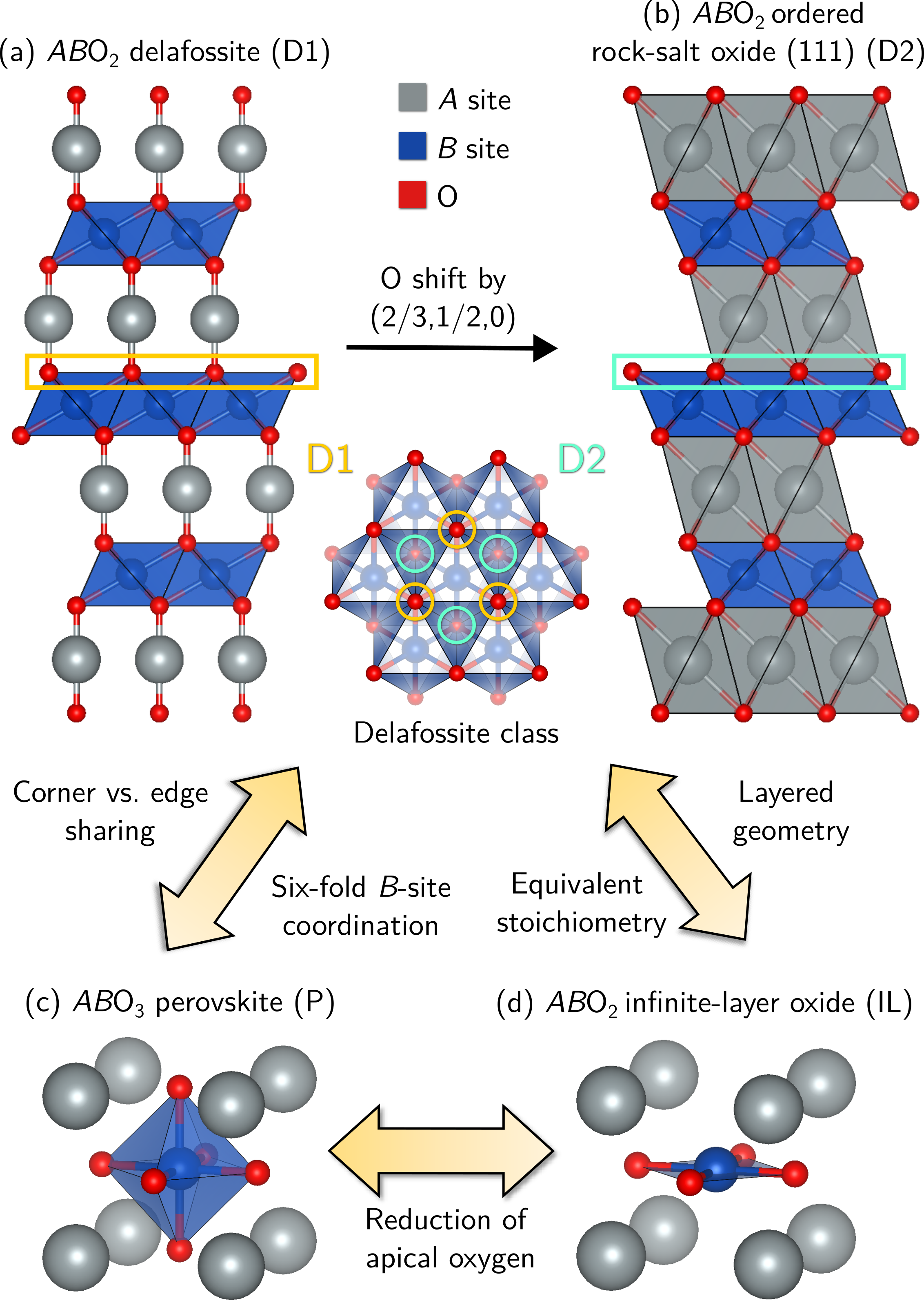}   \\
    \caption{Oxide structures considered here: 
    (a)~The trigonal-layered delafossite (D1, $R\bar{3}m$),
    (b)~its ordered-rock-salt variant with different layer stacking (D2, $R\bar{3}m$),
    (c)~the cubic perovskite structure (P, $Pm\bar{3}m$),
    (d)~and the square-planar infinite-layer geometry (IL, $P4/mmm$).
    In D1, $B$ describes the octahedral site (blue).
    In D2, both the $A$ sites (gray) and the $B$ sites (blue) are octahedrally coordinated;
    thus, we define $B$ as the smaller octahedron. 
    The arrows highlight the interrelations between the different phases.
    }
    \label{fig:Structures}
\end{figure}

An intriguing layered oxide that combines the infinite-layer $AB$O$_2$ stoichiometry
with the perovskite-like octahedral motif is the delafossites (Fig.~\ref{fig:Structures}).
In sharp contrast to the tetragonal infinite-layer geometry, 
delafossites (here referred to as D1) can be considered as the parent compound of various triangular lattices
and are closely related to honeycomb and Kagom\'e systems~\cite{Lechermann-Delafossite:21}.
They present hexagonal $B$O$_2$ 
layers with six-fold oxygen coordination 
that alternate with $A$ planes acting as charge reservoirs [Fig.~\ref{fig:Structures}(a)].
In particular, Pd- and Pt-based delafossites are currently the subject of intense research~\cite{mackenzie_properties_2017},
e.g., due to their quasi-two-dimensional properties and highly anisotropic conductivity and large out-of-plane Seebeck coefficient~\cite{Ong:10, Gruner:15, yordanov_large_2019}. 
Planar shift of the oxygen layer by $(2/3, 1/3, 0)$ 
results in the ordered rock-salt (111) variant, which we denote as D2 [Fig.~\ref{fig:Structures}(b)].

Here we construct a comprehensive and consistent database of layered trigonal $AB$O$_2$ delafossites and ordered rock-salt oxides by high-throughput density functional theory simulations including an on-site Coulomb repulsion term.
In conjunction with earlier data on infinite-layer and perovskite oxides, we assess the relative thermodynamic stability of different competing $AB$O$_2$ structural motifs and screen isoelectronic alternatives to infinite-layer nickelates.
Remarkably, we find that the delafossite structure rivals the infinite-layer phase in thermodynamic stability for the nickelates, and even
more for the palladate and platinate analogs. 
Concomitantly, we find that the latter two show a significantly reduced energy cost for the topotactic reduction from the perovskite phase.
Analysis of the electronic structure of the delafossite nickelates, palladates, and platinates, distinguished by a reversed cation order, reveals a strongly $A$-site $d_{z^2}$-dominated Fermi surface, markedly different from the $d_{x^2-y^2}$ sheets typical of infinite-layer compounds.
We identify the La-Ni combination as thermodynamic optimum for stabilizing the infinite-layer geometry.
Moreover, we demonstrate that Ca, Sr, and Ba hole doping systematically enhances the relative stability of the infinite-layer phase across all three transition-metal families.
These findings highlight fundamental challenges in synthesizing substrate-free infinite-layer oxides
and provide valuable guidance for future experimental efforts aimed at realizing novel superconducting compounds beyond rare-earth nickelates.

\begin{figure*}
    \includegraphics[width=\textwidth]{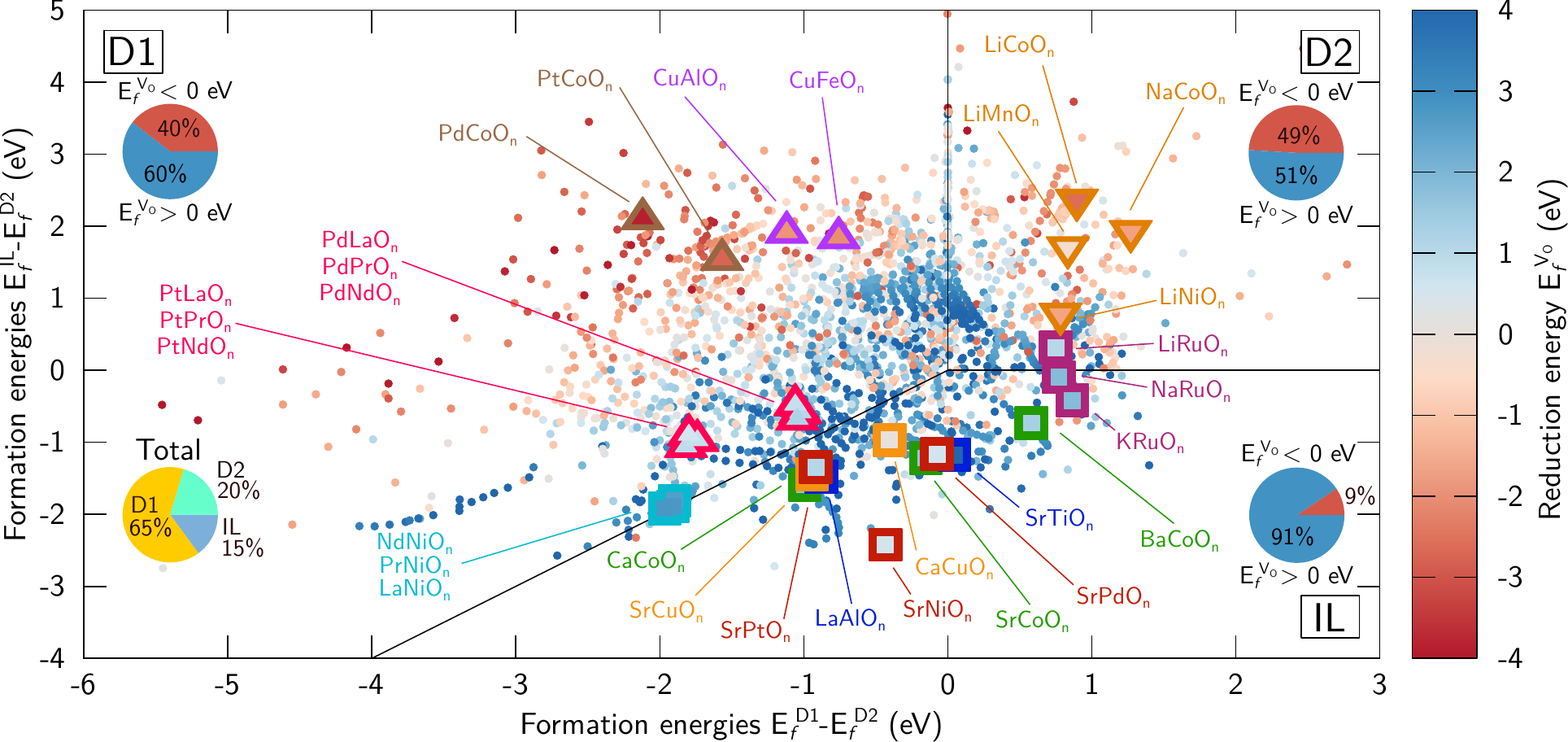}
    \caption{
        Phase diagram obtained from first principles, comparing simultaneously the relative stability of delafossite (D1), ordered rock-salt (D2), and infinite-layer oxides (IL) 
        versus the perovskite phase (P) 
        under oxygen-rich conditions for $2346$ element combinations (data points).
        The point colors encode the energy associated with the reduction reaction from the 
        P phase ($n=3$) to the respective ground state in the 
        D1, D2, IL space ($n=2$), considering also the possibility of reversed $A$-$B$ order (see Table~\ref{tab:Ef}).
        The three $n=2$ phases are represented by the three sectors. 
        In addition to the rare-earth nickelates, palladates, and platinates,
        several prominent compounds are marked for reference. 
        The Sr-based Ni, Pd, and Pt compounds demonstrate trends related to hole doping.
        The individual pie charts in each sector quantify that most element combinations stabilize rather the P phase ($E_f^{V_\text{O}} > 0$, blue),
        whereas the 'total' pie chart shows that D1 is the preferred phase for $n=2$.
        }
    \label{fig:PhaseDiagram}
\end{figure*}

\section{Methods and data generation}

We performed first-principles simulations 
in the framework of density functional theory~\cite{KoSh65} (DFT) using the projector-augmented wave formalism (PAW)
as implemented in the Vienna \textit{ab initio} simulation package~\cite{USPP-PAW:99, PAW:94}
using the generalized gradient approximation 
as parameterized by Perdew, Burke, and Ernzerhof~\cite{PeBu96}
to construct a database of ground-state energies and optimized lattice parameters for $4692$ different element combinations at the $A$ and $B$ sites (as detailed below) in the D1 and D2 geometries [Fig.~\ref{fig:Structures}(a,b)] modeled by using rhombohedral unit cells.
Some D2 compounds spontaneously interchanged their $A$ and $B$ sites during relaxation,
which we found to be suppressed for D1 by a kinetic barrier.
In total, we obtained 7038 unique configurations. 

Consistent with previous work on the IL and P phases~\cite{SahinovicGeisler:21, SahinovicGeisler:22} [Fig.~\ref{fig:Structures}(c,d)],
we used an energy cutoff of $520$~eV, PAW
and $U=3.32$, $3.7$, $5.3$, $3.9$, $4.38$, $6.2$, $3.25$, and $6.2$~eV
for Co, Cr, Fe, Mn, Mo, Ni, V, and W $d$ states, respectively,
adopting the DFT$+U$ standards of the Materials Project database~\cite{MatProj:13, PYMATGEN:13, LiechtensteinAnisimov:95}.
A smearing of 5~mRy using the Methfessel-Paxton~\cite{Methfessel_1989} scheme for $k$-point integration in the Brillouin zone was employed,
and we ensured $k$-point convergence of the total energy within a few meV per unit cell (e.g., $10\times10\times10$ for PdCoO$_2$).
These settings simultaneously render accurate lattice parameters.
The formation energies $E_f$ of the different $AB$O$_n$ compounds ($n=2,3$) are determined from the DFT$+U$ ground-state energies via
\begin{equation*}
    E_f^{AB\text{O}_n} = E(AB\text{O}_n) - E(A~\text{bulk}) - E(B~\text{bulk}) - n \, \mu_\text{O} \, ,
\end{equation*}
where $\mu_\text{O} = \frac{1}{2} E(\text{O$_2$})$ models the oxygen-rich limit.
The well-known overbinding of gas-phase O$_2$ molecules in DFT necessitates a correction of $E(\text{O$_2$})$,
which we performed such as to reproduce the experimental O$_2$ binding energy of $5.16$~eV~\cite{GeislerPentcheva-LNOLAO-Resonances:19, GeislerPentcheva-LCO:20, SahinovicGeisler:21, SahinovicGeisler:22}. 
The oxygen vacancy formation energy associated with the reduction from the $AB$O$_3$ perovskite structure to either of the three considered $AB$O$_2$ phases can be obtained via
\begin{equation*}
    E_f^{V_\text{O}} = E(AB\text{O}_2) - E(AB\text{O}_3)  + \mu_\text{O} =  E_f^{AB\text{O}_2} - E_f^{AB\text{O}_3} \, .
\end{equation*}
All energies are given per formula unit.

\begin{table*}
	\centering
\caption{\label{tab:Ef} Ground-state chemical formulas (printed in $AB$O$_n$ order), formation energies $E_f$ (eV), and $E_f^{V_\text{O}}$ (eV) for different element combinations in the delafossite (D1), ordered rock-salt (D2), and infinite-layer phases (IL) (see also Fig.~\ref{fig:PhaseDiagram}).
In each row, the lowest $n=2$ formation energy is marked in bold. 
In addition, reference values for the cubic perovskite phase (P; $n=3$) are provided. 
}
\begin{ruledtabular}
\begin{tabular}{lccccccccccc}
Elements & \multicolumn{3}{c}{--- Delafossite (D1) ---} &\multicolumn{3}{c}{--- Ordered rock-salt (D2) ---} &\multicolumn{3}{c}{--- Infinite-layer (IL) ---} &\multicolumn{2}{c}{--- Perovskite (P) ---}\\
& Formula & $E_f^\text{D1}$ & $E_f^{V_\text{O}}$ & Formula & $E_f^\text{D2}$ & $E_f^{V_\text{O}}$ & Formula & $E_f^\text{IL}$ & $E_f^{V_\text{O}}$ & Formula & $E_f^\text{P}$\\
\hline
La - Ni & NiLaO$_2$ & $\mathbf{-10.46}$ & $2.77$ & LaNiO$_2$ & $-8.49$ & $4.74$ & LaNiO$_2$ & $-10.41$ & $2.82$ & LaNiO$_3$ & $-13.22$ \\
Pr - Ni & NiPrO$_2$ & $\mathbf{-10.09}$ & $2.74$ & PrNiO$_2$ & $-8.19$ & $4.64$ & PrNiO$_2$ & $-10.07$ & $2.76$ & PrNiO$_3$ & $-12.83$ \\
Nd - Ni & NiNdO$_2$ & $\mathbf{-10.22}$ & $2.60$ & NdNiO$_2$ & $-8.32$ & $4.50$ & NdNiO$_2$ & $-10.13$ & $2.69$ & NdNiO$_3$ & $-12.82$ \\
La - Pd & PdLaO$_2$ & $\mathbf{-9.56}$ & $1.20$ & PdLaO$_2$ & $-8.52$ & $2.25$ & LaPdO$_2$ & $-9.20$ & $1.57$ & LaPdO$_3$ & $-10.76$ \\
Pr - Pd & PdPrO$_2$ & $\mathbf{-9.17}$ & $1.03$ & PdPrO$_2$ & $-8.14$ & $2.06$ & PrPdO$_2$ & $-8.73$ & $1.47$ & PrPdO$_3$ & $-10.19$ \\
Nd - Pd & PdNdO$_2$ & $\mathbf{-9.30}$ & $0.79$ & PdNdO$_2$ & $-8.24$ & $1.85$ & NdPdO$_2$ & $-8.72$ & $1.37$ & NdPdO$_3$ & $-10.09$ \\
La - Pt & PtLaO$_2$ & $\mathbf{-9.17}$  & $0.91$ & LaPtO$_2$ & $-7.35$  & $2.73$ & LaPtO$_2$ & $-8.41$ & $1.67$ & LaPtO$_3$ & $-10.08$ \\
Pr - Pt & PtPrO$_2$ & $\mathbf{-8.70}$  & $0.76$ & PrPtO$_2$ & $-6.96$  & $2.50$ & PrPtO$_2$ & $-7.94$ & $1.52$ & PrPtO$_3$ & $-9.46$ \\
Nd - Pt & PtNdO$_2$ & $\mathbf{-8.82}$  & $0.52$ & NdPtO$_2$ & $-7.02$  & $2.32$ & NdPtO$_2$ & $-8.41$ & $0.93$ & NdPtO$_3$ & $-9.34$ \\
\hline
Sr - Ni & NiSrO$_2$ & $-7.40$  & $2.48$ & SrNiO$_2$ & $-6.96$  & $2.92$ & SrNiO$_2$ & $\mathbf{-9.38}$ & $0.50$ & SrNiO$_3$ & $-9.88$ \\
Sr - Pd & PdSrO$_2$ & $-6.67$  & $1.63$ & SrPdO$_2$ & $-6.59$  & $1.71$ & SrPdO$_2$ & $\mathbf{-7.76}$ & $0.54$ & SrPdO$_3$ & $-8.30$ \\
Sr - Pt & PtSrO$_2$ & $-6.91$  & $1.49$ & SrPtO$_2$ & $-5.99$  & $2.41$ & SrPtO$_2$ & $\mathbf{-7.34}$ & $1.06$ & SrPtO$_3$ & $-8.40$ \\
\hline
Ca - Co & CoCaO$_2$ & $-7.37$  & $2.63$ & CaCoO$_2$ & $-7.95$  & $2.05$ & CaCoO$_2$ & $\mathbf{-8.69}$ & $1.31$ & CaCoO$_3$ & $-10.00$ \\
Sr - Co & CoSrO$_2$ & $-7.29$  & $2.76$ & SrCoO$_2$ & $-7.14$  & $2.91$ & SrCoO$_2$ & $\mathbf{-8.35}$ & $1.30$ & SrCoO$_3$& $-10.05$ \\
Ba - Co & CoBaO$_2$ & $-7.14$  & $2.33$ & BaCoO$_2$ & $-6.15$  & $3.32$ & BaCoO$_2$ & $\mathbf{-8.35}$ & $1.12$ & BaCoO$_3$ & $-9.47$ \\
Pd - Co & PdCoO$_2$ & $\mathbf{-5.17}$ & $-2.78$ & PdCoO$_2$ & $-3.61$ & $-1.21$ & PdCoO$_2$ & $-0.94$ & $1.46$ & PdCoO$_3$ & $-2.39$  \\
Pt - Co & PtCoO$_2$ & $\mathbf{-5.16}$ & $-3.88$ & PtCoO$_2$ & $-3.05$ & $-1.77$ & PtCoO$_2$ & $-0.94$ & $0.34$ & PtCoO$_3$ & $-1.28$  \\
Na - Co & NaCoO$_2$ & $-5.92$ & $-0.49$ & NaCoO$_2$ & $\mathbf{-7.19}$ & $-1.76$ & NaCoO$_2$ & $-5.26$ & $0.16$ & NaCoO$_3$ & $-5.42$  \\
\hline
Ca - Cu & CuCaO$_2$ & $-7.94$  & $0.55$ & CaCuO$_2$ & $-7.53$  & $0.96$ & CaCuO$_2$ & $\mathbf{-8.50}$ & $-0.01$ & CaCuO$_3$ & $-8.49$ \\
Sr - Cu & CuSrO$_2$ & $-7.74$  & $0.94$ & SrCuO$_2$ & $-6.80$  & $1.99$ & SrCuO$_2$ & $\mathbf{-8.24}$ & $0.44$ & SrCuO$_3$ & $-8.68$ \\
Al - Cu & CuAlO$_2$ & $\mathbf{-9.66}$ & $-1.99$ & CuAlO$_2$ & $-8.54$ & $-0.87$ & CuAlO$_2$ & $-6.62$ & $1.05$ & CuAlO$_3$ & $-7.67$ \\
Fe - Cu & CuFeO$_2$ & $\mathbf{-5.49}$ & $-2.10$ & CuFeO$_2$ & $-4.73$ & $-1.34$ & CuFeO$_2$ & $-2.90$ & $0.50$ & CuFeO$_3$ & $-3.39$ \\
\end{tabular}
\end{ruledtabular}
\end{table*}

\section{Results and discussion}

\subsection{Data exploration and phase diagram: Delafossite versus infinite-layer structure for nickelates, palladates, and platinates}

The large and consistent first-principles dataset generated here
allows us to obtain unprecedented insight into the relative stability of the four considered oxide classes across the periodic table.
We compile a phase diagram from the D1, D2, IL, and P formation energies
that illustrates their relative stability in a compact form (Fig.~\ref{fig:PhaseDiagram}).
We find that the ordering of the $A$ and $B$ cations can vary across different structural phases for a fixed elemental combination.
The phase diagram is constructed solely from the lowest-energy configuration (ground state) for each combination,
resulting in a total of $2346 = 4692/2$ unique data points.
In conjunction with previously obtained IL and P data~\cite{SahinovicGeisler:21, SahinovicGeisler:22},
our analysis is based on $16{,}422$ individual DFT structure optimizations.

The horizontal axis in Fig.~\ref{fig:PhaseDiagram} corresponds to $E_f^\text{D1} - E_f^\text{D2}$,
and the data ranges from $\sim -6$ to $3$~eV,
while $E_f^\text{IL} - E_f^\text{D2}$ (vertical axis)
ranges from $\sim -4$ to $5$~eV.
The black lines correspond to
$E_f^\text{D1} - E_f^\text{IL} = 0$,
$E_f^\text{D1} - E_f^\text{D2} = 0$, and
$E_f^\text{IL} - E_f^\text{D2} = 0$ and
partition the data into three sectors of thermodynamic stability: D1, D2, and IL.
As a benchmark, we highlight several well-known materials, namely CuAlO$_2$, CuFeO$_2$, PdCoO$_2$, PtCoO$_2$, and NaCoO$_2$, all of which are correctly predicted to adopt their experimentally observed ground-state phases.
Moreover, the computed negative reduction energies confirm that the $AB$O$_2$ stoichiometry (red) is thermodynamically favored over the $AB$O$_3$ P phase (blue).
Additionally, we identify the widely studied battery cathode materials LiCoO$_2$, LiMnO$_2$, and LiNiO$_2$ \cite{Battery_materials_06, WANG2020155827} within the D2 regime, also exhibiting negative reduction energies. Interestingly, topical alkali ruthenates such as NaRuO$_2$ \cite{Ortiz_NaRuO2_2023} are found near the D2-IL phase boundary. In contrast, alkaline-earth cobaltates, including CaCoO$_2$ \cite{Kim-CaCoO2-IL:23}, are clearly located within the IL regime. 
For selected compounds, the computed reduction energies are explicitly reported in Table~\ref{tab:Ef}.

The 'total' pie chart in Fig.~\ref{fig:PhaseDiagram} reflects the number of data points in the different sectors.
We observe a strong preference for D1 (65\%) and D2 (20\%) over the IL phase, which only hosts 15\% of the data points.
Moreover, only 9\% of the data points in the IL regime exhibit negative reduction energies, whereas 91\% prefer the P phase.
This demonstrates that the IL phase is rather exotic, and that the trigonal D1 and D2 geometries are statistically preferred.
Surprisingly, further analysis of these 9\% uncovers seven IL compounds below or on the convex hull constructed using the Materials Project database,
which are summarized in Table~\ref{tab:ILConvexHull}.
Among these seven compounds, exclusively BaAgO$_2$ is isoelectronic to the IL cuprates
and presents a highly similar band structure (see Appendix). 
Although Ag compounds (e.g., AgF$_2$) have been considered as potential cuprate analogs~\cite{AgF2:22},
BaAgO$_2$ has so far been explored in different contexts, such as localized surface plasmon resonance for heat shielding applications~\cite{BaAgO2:21}, and should be revisited for superconductivity.

\begin{table}[b]
	\centering
    \caption{\label{tab:ILConvexHull}Infinite-layer oxides identified below and on the convex hull ($E_\text{hull}$ denotes the energy distance).
    BaAgO$_2$ is isoelectronic to the infinite-layer cuprates.}
\begin{ruledtabular}
\begin{tabular}{lccc}
Material & $E_\text{hull}$ (eV/atom) & $a$ (\AA) & $c$ (\AA) \\
\hline
RbIrO$_2$ & $-0.110$ & $3.86$ & $4.77$ \\
RbPtO$_2$ & $-0.087$ & $3.95$ & $4.57$ \\
CsRhO$_2$ & $-0.070$ & $3.89$ & $5.05$ \\
CsPtO$_2$ & $-0.048$ & $3.95$ & $4.99$ \\
CsOsO$_2$ & $-0.013$ & $3.82$ & $5.35$ \\
RbOsO$_2$ & $-0.008$ & $3.82$ & $4.96$ \\
BaAgO$_2$ & $0.0$    & $4.34$ & $3.74$ \\
\end{tabular}
\end{ruledtabular}
\end{table}

We now analyze the phase diagram in Fig.~\ref{fig:PhaseDiagram} with particular emphasis on the trends observed for nickelates, palladates, and platinates.
The cuprates CaCuO$_2$ and SrCuO$_2$ are predicted to lie in the IL regime. 
In sharp contrast,
the rare-earth nickelates $R^{3+}$NiO$_2$, 
despite being formally isoelectronic to the cuprates,
can be identified close to the D1-IL phase boundary.
Thus, the delafossite structure emerges as a strong competitor to the IL phase for the nickelates;
Table~\ref{tab:Ef} even reveals a minor energetic preference of $0.02$-$0.09$~eV
for the D1 structure. 
Interestingly, this trend is strongly enhanced for the palladates and platinates,
where the D1 phase is $\sim 0.3$-$0.8$~eV more stable than the IL phase 
for the La, Pr, and Nd compounds.

Secondly, we observe that the $A$- and $B$-site elements are \textit{reversed} in the D1 phase,
resulting in (Ni,Pd,Pt)$R$O$_2$,
whereas in the IL phase, the rare-earth ion preferably occupies the $A$ site (Table~\ref{tab:Ef}).
We explain this phenomenon with the most prevalent delafossite oxidation state $A^{1+}B^{3+}$O$_2$~\cite{mackenzie_properties_2017},
according to which a Ni$^{1+}$, Pd$^{1+}$, or Pt$^{1+}$ ion prefers the $A$ site, whereas the $R^{3+}$ ion occupies the $B$ site.
The implications for the electronic structure are investigated below.

Thirdly, we note that nickelates, palladates, and platinates generally feature a positive $E_f^{V_\text{O}}$, which implies that the P phase is more stable under oxygen-rich conditions.
For the palladates and platinates, we find $E_f^{V_\text{O}} \sim 0.52$-$1.20$~eV, which is $1.5$~eV lower than for the nickelates.

In conclusion, the assessment of Fig.~\ref{fig:PhaseDiagram} and Table~\ref{tab:Ef} establishes
that the growth of IL rare-earth nickelates, palladates, and platinates
is impeded by at least three fundamental challenges: the energetic preference for the D1 phase, reversed $A$- and $B$-site order, as well as a positive oxygen vacancy formation energy.
These results explain the importance of enforcing the tetragonal symmetry 
to stabilize the IL geometry, e.g., by epitaxial growth on a substrate.
We emphasize that this is particularly relevant for the palladates and platinates.

Intriguingly, we find that the Ca-, Sr-, and Ba-based compounds consistently favor the IL phase as their ground state (Fig.~\ref{fig:PhaseDiagram}, Table~\ref{tab:Ef}; see also Appendix).
We speculate that even partial Ca, Sr, Ba substitution, as commonly employed to hole-dope IL rare-earth nickelates into the superconducting dome, may play a crucial role in stabilizing the IL phase over the competing D1 and D2 structures.
Furthermore, we see that such substitution tends to lower the reduction energy relative to the parent P phase (Table~\ref{tab:Ef}),
potentially facilitating the topotactic reduction process.

\subsection{Stability trends for delafossite, ordered rock-salt, and infinite-layer oxides}

\begin{figure}[!htbp]
    \includegraphics[width=\linewidth]{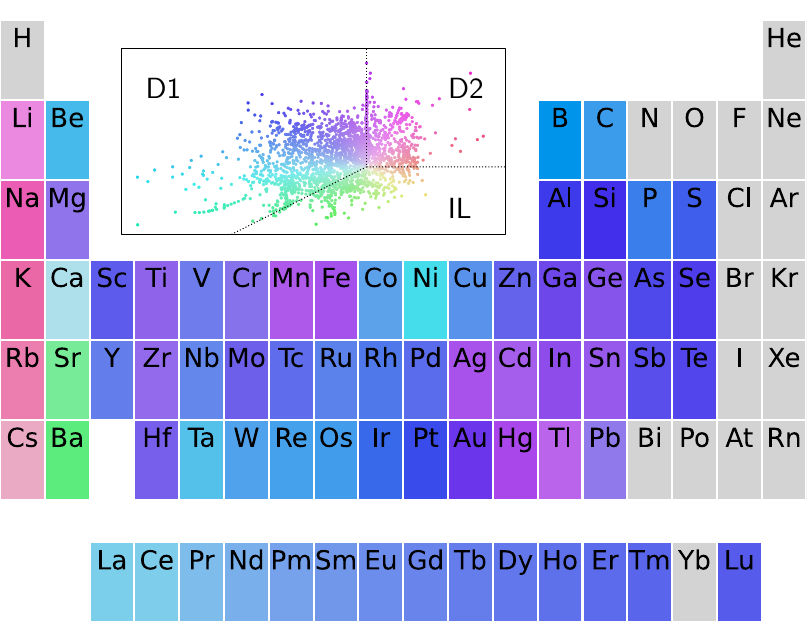}
    \caption{Stability trends of the D1, D2, and IL oxide phases across the periodic table.
    Each element is colored according to Eqs.~\ref{eq:radius_phi_conversion} 
    and highlights which phase will statistically stabilize
    if this element is involved in the compound at either site $A$ or $B$.
    The inset clarifies the color assignment explicitly for each data point in the phase diagram (Fig.~\ref{fig:PhaseDiagram}):
    The D1 materials can be found between cyan and purple,
    the D2 materials between purple an red,
    and the IL materials between yellow and green.
    Elements depicted in grey are not considered here.
    }
    \label{fig:PSE}
\end{figure}

Next, we determine the statistical trends that link the relative stability of the three different $n=2$ oxides
and the involved chemical elements $Z$ by constructing a compact periodic table diagram from the data (Fig.~\ref{fig:PSE}). 
In turn, this provides design guidelines on how to stabilize each of the desired phases.
To this extent, the positions of the 68 compositions in the phase diagram 
involving a given chemical element $Z$ at either $A$ or $B$ site
are first averaged, and the result is subsequently converted into polar coordinates:
\begin{equation}
\begin{split}
    \label{eq:radius_phi_conversion}
    r(Z) &= \sqrt{\langle E_f^\text{IL}-E_f^\text{D2}\rangle_Z^2+\langle E_f^\text{D1}-E_f^\text{D2} \rangle_Z^2} \, \\
    \varphi(Z) &=  \text{atan2}(\langle E_f^\text{IL}-E_f^\text{D2}\rangle_Z,\langle E_f^\text{D1}-E_f^\text{D2}\rangle_Z) \ . \\
\end{split}
\end{equation}
Here, $r=0$ corresponds to the intersection of the three stability sectors. 
Subsequently, $r$ and $\varphi$ are visualized by mapping to the saturation and hue of the color wheel, respectively.
In this representation, the color reflects the stability sector in which a given composition is located,
whereas the saturation indicates its distance from the origin.
The inset in Fig.~\ref{fig:PSE} demonstrates this color assignment explicitly for each data point in the phase diagram. 
Finally, all elements $Z$ in the periodic table are colored accordingly.

Figure~\ref{fig:PSE} shows that an involvement of alkali metals promotes formation of the D2 phase (red color).  
The alkaline earths exhibit a transition from D1 (Be, Mg; blue-purple) to IL (Ca, Sr, Ba; green),
in line with our above conclusion that Ca, Sr, Ba hole doping tends to stabilize the IL structure.
The rare-earth elements clearly stabilize the D1 structure, and this trend is continuously enhanced throughout the series (light blue to dark blue). 
This implies that the IL phase is significantly easier to fabricate for the early rare-earth metals,
whereas detrimental competition with the D1 phase increases for the later rare-earth metals.

Many prominent delafossites combine transition metals at both the $A$ and $B$ sites,
such as PdCoO$_2$, CuFeO$_2$, and AgNiO$_2$.
We corroborate this so-far empirical observation in Figure~\ref{fig:PSE},
which shows that the majority of the transition metals drive the stabilization of the D1 phase. 
Interestingly, we observe a strong trend towards D2 (purple) around Mn, Fe ($3d$), Ag, Cd ($4d$) and Hg, Tl ($5d$),
forming a quasi-diagonal pattern in the periodic table.

Surprisingly, Ni emerges as a pronounced sweet spot with a strong tendency towards the IL phase (light blue),
in sharp contrast to Pd and Pt, which statistically prefer the D1 phase (dark blue).
This is consistent with the conclusions drawn above.
Again, a diagonal pattern in the periodic table can be identified:
The light-blue color around Co, Ni, Cu ($3d$) shifts roughly via Ru ($4d$) to Ta-Os ($5d$) and subsequently to the early rare-earth metals.
Our statistical analysis highlights La-Ni as the most favorable rare-earth/transition-metal pairing for stabilizing the IL phase within the Ni, Pd, and Pt oxides.

\subsection{Electronic structure of delafossite versus infinite-layer nickelates, palladates, and platinates} 

\begin{figure}[!htbp]
    \includegraphics[width=0.99\linewidth]{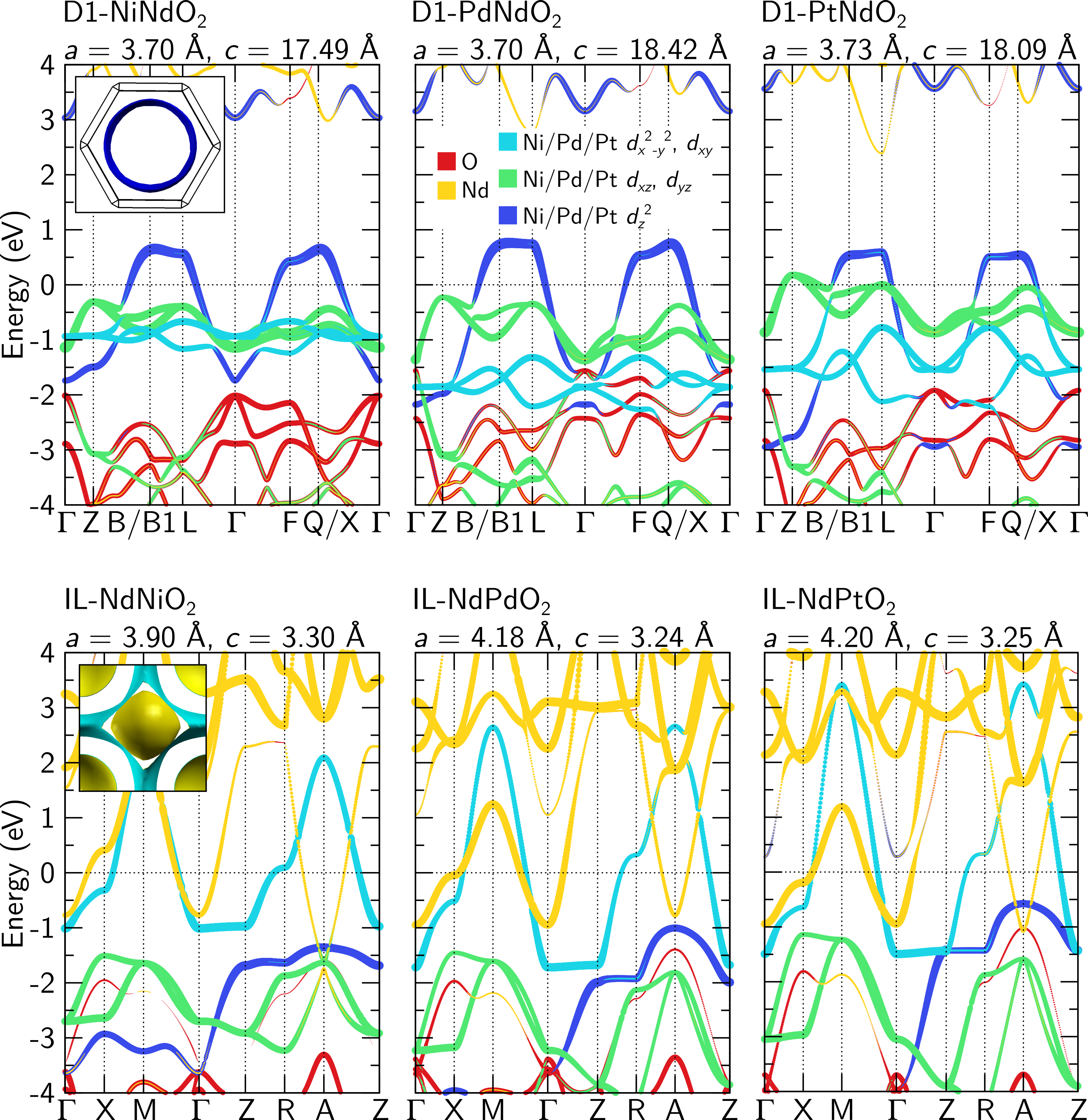}
    \caption{Orbital-resolved band structure of delafossite (D1, top row) versus infinite-layer (IL, bottom row) neodymium nickelates, palladates, and platinates (left to right).
    The transition-metal $d$ bands corresponding to the $d_{z^2}$, ($d_{x^2-y^2}$, $d_{xy}$), and ($d_{xz}$, $d_{yz}$) representations
    in the rhombohedral delafossites (transition metal at the $A$ site) are marked in dark blue, cyan, and green, respectively.
    In the tetragonal infinite-layer case (transition metal at the $B$ site),
    the $d_{x^2-y^2}$- and $d_{xy}$-derived bands are further split, the latter being below the shown energy range. 
    The oxygen and Nd contributions are displayed in red and yellow, respectively.
    The insets compare the Fermi surfaces of D1 versus IL for the nickelate compound, viewed along the $z$~axis.
    Static correlation effects are considered by employing $U=4$, $3$, and $2$~eV at the Ni, Pd, and Pt sites, respectively.
    }
    \label{fig:BS_NiPdPt}
\end{figure}

The analysis above established D1 as a strong competitor to the IL phase for the rare-earth nickelates
and particularly for the palladates and platinates.
Moreover, we found that in the D1 structure the rare-earth ion occupies the octahedrally coordinated $B$ site,
which may promote novel quantum properties such as magnetic frustration~\cite{hashimoto_structures_2002, miyasaka_synthesis_2009, Alkali-Rare-Earth-D2}.
Simultaneously, the transition-metal ion at the $A$ site adopts a rather peculiar two-fold coordination.

In order to elucidate the key electronic differences between the D1 and IL phases,
Fig.~\ref{fig:BS_NiPdPt} compares the orbital-resolved band structure
for the neodymium nickelates, palladates, and platinates.
We employ $U=4$ at the Ni sites, consistent with earlier electronic structure work \cite{GeislerPentcheva-NNOCCOSTO:21},
and lower values of $3$ and $2$~eV for Pd and Pt, respectively.
We see that the transition-metal states are always the active states at the Fermi level,
irrespective of the fundamentally different chemical environments in the D1 versus IL geometry.
In sharp contrast, the rare-earth ion,
which contributes electron pockets around the $\Gamma$ and $A$ points in the IL phase~\cite{Botana-Inf-Nickelates:19, Sakakibara:20, Lechermann-Inf:20},
is found to be completely inert in the D1 structure, where the respective states can be observed above $\sim 3$~eV.

The crystal field experienced by the transition-metal $d$ states at the linearly coordinated $A$ site in the D1 structure
splits them in the groupings of $d_{z^2}$, ($d_{x^2-y^2}$, $d_{xy}$), and ($d_{xz}$, $d_{yz}$) (Fig.~\ref{fig:BS_NiPdPt}). 
For NiNdO$_2$ and PdNdO$_2$, exclusively the $d_{z^2}$ states contribute to the Fermi surface
and exhibit a particularly high dispersion along $Z$-$B$, $L$-$\Gamma$, $\Gamma$-$F$, and $X$-$\Gamma$.
The $d_{xz}$ and $d_{yz}$ orbitals, which can be identified directly below the Fermi level in NiNdO$_2$ and PdNdO$_2$,
slightly cross the Fermi level for PtNdO$_2$ and create an additional hole pocket around the $Z$~point.
For PdNdO$_2$ and PtNdO$_2$,
the $d_{x^2-y^2}$ and $d_{xy}$ states are located at even lower energies close to the oxygen states,
whereas for NiNdO$_2$, the ($d_{x^2-y^2}$, $d_{xy}$) and ($d_{xz}$, $d_{yz}$) states exhibit comparable energies.

While the overall shape of the band structure is similar among the three D1 and IL systems,
it is highly different between D1 versus IL.
The distinct chemical environment in the IL phase lifts the degeneracy of the $d_{x^2-y^2}$ and $d_{xy}$ states (Fig.~\ref{fig:BS_NiPdPt}). 
Now, the $d_{x^2-y^2}$-derived bands exhibit the highest dispersion among the transition-metal states,
which we find to be continuously enhanced from NdNiO$_2$ to NdPtO$_2$.
In contrast to the D1 phase, the $d_{z^2}$ states are fully occupied and found at energies similar to those of the $d_{xz}$ and $d_{yz}$ states.
Hence, we find that these two orbitals reverse their fundamental role in delafossite versus infinite-layer nickelates, palladates, and platinates.

\section{Summary}

We presented a comprehensive high-throughput study of $AB$O$_2$ oxides, generating an extensive database of 7038 unique delafossite and ordered rock-salt (111) compounds by performing density functional theory simulations including an on-site Coulomb repulsion term.
In conjunction with previous data of 9384 perovskite and infinite-layer materials,
we constructed a detailed phase diagram and systematically assessed the relative thermodynamic stability
of these four distinct oxides across the periodic table.
The results revealed that the delafossite geometry, which intriguingly combines the infinite-layer stoichiometry with perovskite-like octahedral $B$-site coordination, emerges as a strong and previously overlooked competitor to the infinite-layer phase.
Our statistical analysis identified the infinite-layer structure as relatively rare but distinct, exhibiting the lowest formation energy for only $15\% \times 9\% = 1.35\%$ of the sampled compounds.
We uncovered La-Ni as the most favorable rare-earth/transition-metal pairing for stabilizing the infinite-layer phase.
In contrast, the isoelectronic Pd and Pt compounds face intrinsic thermodynamic barriers to adopting this structure, yet show significantly reduced oxygen reduction energies from the perovskite parent phase by $\sim 1.5$~eV.
Analysis of the electronic structure demonstrated that the respective neodymium-based delafossites exhibit $d_{z^2}$-dominated Fermi surfaces, as opposed to the strong $d_{x^2-y^2}$ character typical of the superconducting infinite-layer nickelates.
Furthermore, we highlighted the critical role of alkaline-earth substitution via Ca, Sr, and Ba, which we found to stabilize the infinite-layer phase for the Ni, Pd, and Pt systems.
Interestingly, introduction of these elements not only shifts the thermodynamic balance towards tetragonal infinite-layer symmetry, but also further lowers the reduction energy, facilitating synthesis via thin-film growth.
This mechanism favorably coincides with their known role as hole dopants in superconducting nickelates and may promote the viability of Ca, Sr, and Ba-substituted Pd and Pt compounds as potential infinite-layer analogs.
Finally, convex hull analysis identified BaAgO$_2$ as a stable and promising isoelectronic alternative to infinite-layer nickelates.
These findings highlight the competition between distinct structural motifs in the $AB$O$_2$ family and provide quantitative guidance to tailor stability through chemical substitution. They also uncover key challenges for synthesizing free-standing infinite-layer oxides and propose strategies to realize new superconductors beyond rare-earth nickelates.

\begin{acknowledgments}
This work was supported by the German Research Foundation (Deutsche Forschungsgemeinschaft, DFG) 
within IRTG 2803 (Projektnummer 461605777),
and the National Science Foundation, Grants No.~NSF-DMR-2118718 and No.~NSF-DMREF-2522891.
Computing time was granted by the Center for Computational Sciences and Simulation of the University of Duisburg-Essen
(DFG Grants No.~INST 20876/209-1 FUGG and No.~INST 20876/243-1 FUGG).
B.G.\ acknowledges start-up funding provided by
the Excellent Early Career Researchers Funding Competition of the University of Duisburg-Essen
as well as the Department of Physics of the University of Duisburg-Essen.
\end{acknowledgments}

%
%

\appendix

\section*{Appendix A: Explicit formation energies for nickelates, palladates, platinates, and cuprates}

\begin{figure}[h]
    \includegraphics[width=\linewidth]{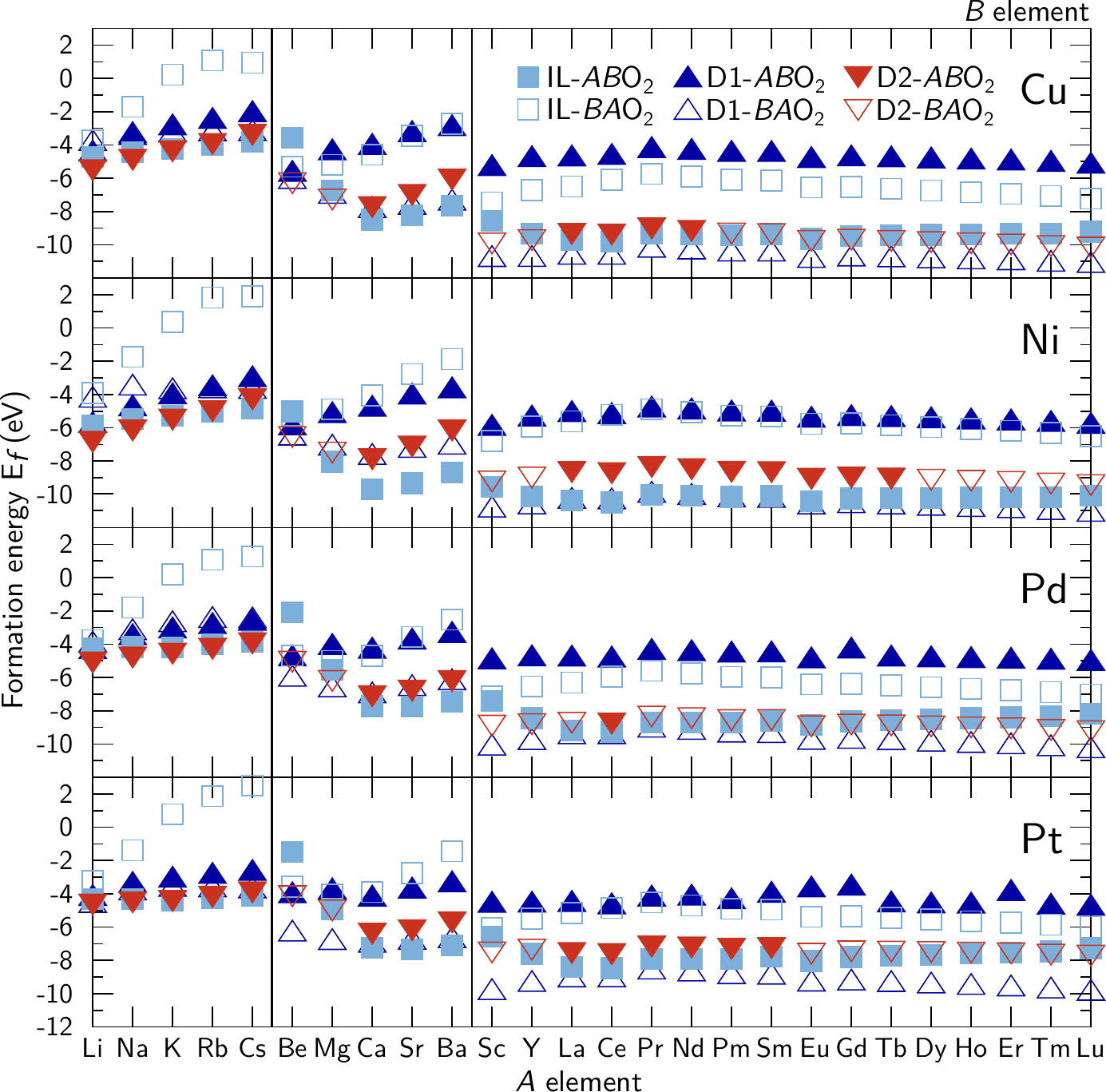}
    \caption{Formation energies $E_f^{}$ of the D1, D2, and IL phases for
        cuprates, nickelates, palladates, and platinates
        as a function of element~$A$. 
        In addition to the filled symbols representing $AB$O$_2$ order,
        open symbols correspond to reversed $BA$O$_2$ order for the D1 and IL phases.
        The D2 geometry always relaxes to its $AB$O$_2$ versus $BA$O$_2$ ground state due to the absence of a kinetic barrier
        and is thus represented by a single symbol (either filled or open).
        }
    \label{fig:E-1}
    \vspace{-1em}
\end{figure}

Figure~\ref{fig:E-1} provides a complementary perspective on the relative stability of the D1, D2, and IL phases
specifically for $B=$~Cu, Ni, Pd, and Pt,
varying the $A$-site element across groups 1, 2 and 3 including rare-earths metals
Moreover, this plot provides the energy related to $A$-$B$ interchange.
Overall, the formation energy curves exhibit similar trends:
The ground-state energy lowers progressively while shifting from group 1 to group 3.

For the alkali metals, we universally observe a close competition between the D2 and IL phase in $AB$O$_2$ order for all considered $B$-site elements.
In contrast, the least stable phase is IL in reversed $BA$O$_2$ order.

Among the alkaline earths, $A=$~Ca, Sr, and Ba stabilize an IL ground state.
This holds particularly in combination with Ni, where also $A=$~Mg results in an IL ground state.
The closest competing phase is generally D1 with reversed $BA$O$_2$ order;
the associated energy differences for the Sr - Ni, Sr - Pd, and Sr - Pt compounds are $1.98$, $1.09$, and $0.43$~eV, respectively (Table~\ref{tab:Ef}).
For Pd-based compounds, Ba doping emerges as a superior alternative to Ca and Sr doping due to an enhanced IL-D1 energy difference.
The IL phase is found to be highly robust against the formation of anti-site defects,
which are impeded by a very high interchange energy (e.g., $\sim 6.6$~eV for Sr - Ni).

In combination with either the group-3 elements Y, Sc or the rare-earth elements La-Lu,
the reversed D1 phase exhibits universally the lowest energy.
The closest competing phases are IL ($AB$O$_2$ order) and D2 (either $AB$O$_2$ or $BA$O$_2$ order).
Interestingly, Fig.~\ref{fig:E-1} explicitly shows that the D1-IL energy difference is minimal for the early rare-earth metals,
but increases significantly for the late rare-earth metals and for Sc and Y.
Moreover, the D1-IL energy differences are considerably smaller for Ni than for Cu, Pd, and Pt
(e.g., Nd - Ni $0.09$~eV, Nd - Pr $0.41$~eV; Table~\ref{tab:Ef}).
These two insights agree with our earlier conclusion 
that La and Ni constitute the optimal combination to realize a rare-earth-based Ni-group IL compound.
The least stable phases are D1 in $AB$O$_2$ order and IL in reversed $BA$O$_2$ order;
for instance, the respective energy difference for Nd - Ni is $\sim 5.2$~eV.

\section*{Appendix B: Electronic structure of \texorpdfstring{B\lowercase{a}A\lowercase{g}O$_2$}{BaAgO$_2$}}

\begin{figure}[b]
    \includegraphics[width=\linewidth]{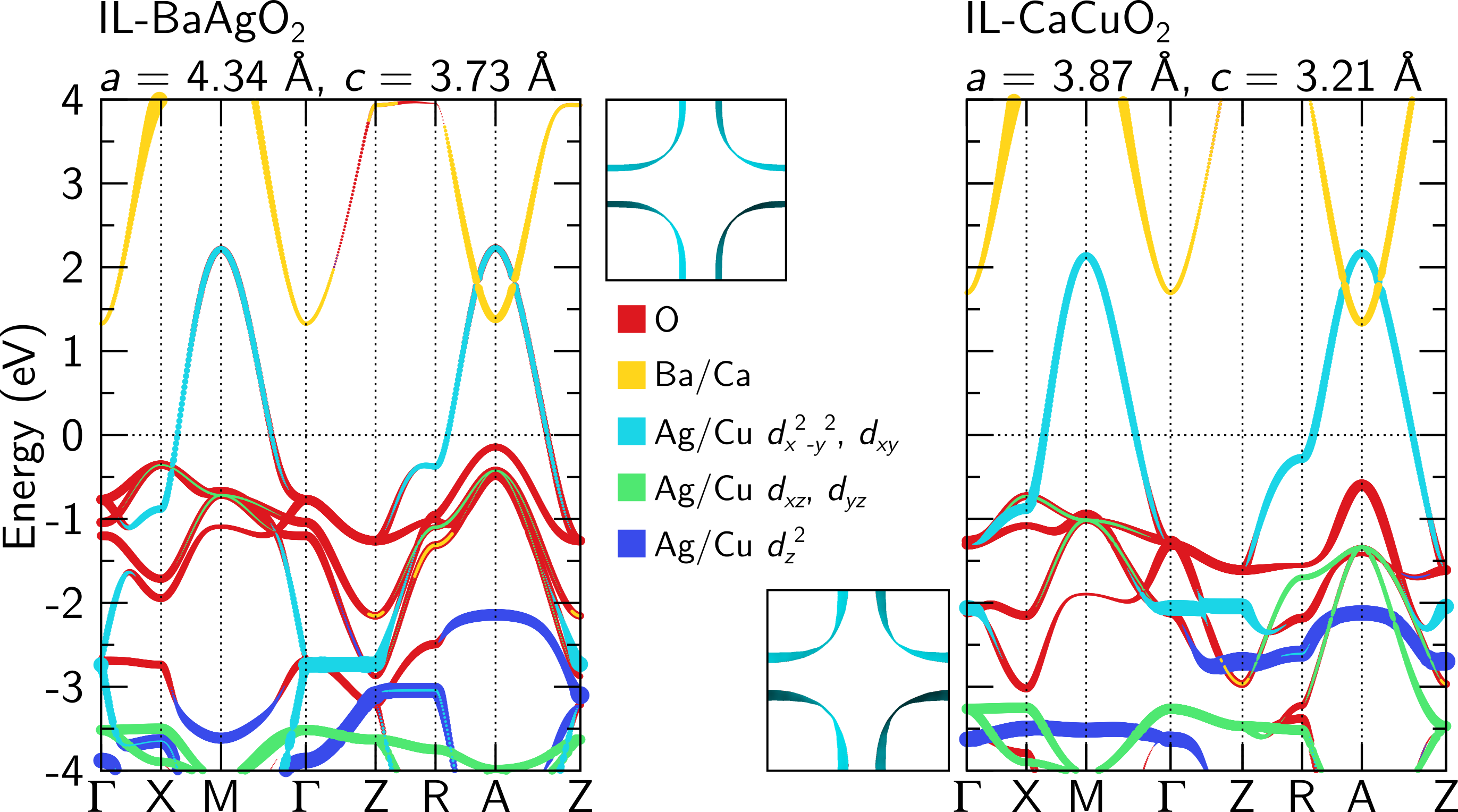}
    \caption{Orbital-resolved band structure and corresponding Fermi surfaces of BaAgO$_2$ (left) and CaCuO$_2$ (right) in infinite-layer geometry, showing the high electronic similarity of the two compounds.
    Consistent with Fig.~\ref{fig:BS_NiPdPt}, the transition-metal $d$ bands are grouped into $d_{x^2-y^2}$ and $d_{xy}$ (cyan), $d_{xz}$ and $d_{yz}$ (green), and $d_{z^2}$ (dark blue).
    The oxygen and Ba/Ca contributions are displayed in red and yellow, respectively.
    For Cu, we apply $U = 4$~eV within DFT$+U$.
    }
    \label{fig:BaAgO2}
\end{figure}

Our high-throughput approach in conjunction with a convex-hull analysis revealed several promising, unexplored stable materials. 
In particular, BaAgO$_2$ emerged as a stable isoelectronic compound to CaCuO$_2$.
Figure~\ref{fig:BaAgO2} compares the orbital-resolved electronic structure of BaAgO$_2$ and CaCuO$_2$,
including a Coulomb repulsion term with $U=4$~eV at the Cu site~\cite{GeislerPentcheva-NNOCCOSTO:21}.  
Both Fermi surfaces are predominantly of $d_{x^2-y^2}$ character and exhibit a highly similar shape.
The characteristic $A$-site (e.g., rare-earth) $d_{z^2}$ hole pockets 
present in the infinite-layer nickelates (see Fig.~\ref{fig:BS_NiPdPt}) are fully depleted here,
and the respective band is found far above the Fermi level at $> 1.3$~eV. 
Concomitantly, states with pronounced oxygen character are observed closely below the Fermi level,
particularly for BaAgO$_2$.
The highly similar electronic structure suggests BaAgO$_2$ as an interesting and potentially superconducting cuprate analog to explore in future work.

\bibliography{BibTeX/sources-msc, BibTeX/sources-bsc, BibTeX/References}

\end{document}